\documentclass[a4paper,11pt]{article}
\pdfoutput=1 % if your are submitting a pdflatex (i.e. if you have
\usepackage{jinstpub} % for details on the use of the package, please
\usepackage{lineno}
\usepackage{siunitx}
\usepackage{xspace}
\usepackage{enumitem}
\setlist{nosep}

\newcommand{\gbps}{~\si[per-mode=symbol]{\giga\bit\per\second}\xspace}
\newcommand{\tbps}{~\si[per-mode=symbol]{\tera\bit\per\second}\xspace}

\newcommand{\ghz}{~\si{\giga\hertz}\xspace}

\newcommand{\mm}{~\si{\milli\meter}\xspace}

\newcommand{\watt}{~\si{\watt}\xspace}

\newcommand{\ps}{~\si{\pico\second}\xspace}
\newcommand{\aunit}[1]{~\si{#1}}
\title{\boldmath PCIe400 generic readout board qualification test }

\author[b]{K.~Arnaud}
\author[c]{A.~Back}
\author[c]{D.~Charlet}
\author[e]{G.~Degret}
\author[e]{L.~Del~Buono}
\author[a]{P.~Durante}
\author[f]{A.~Hervo}
\author[b]{F.~Hachon}
\author[c]{X.~Lafay}
\author[b,1]{J.~Langou{\"e}t\note{Corresponding author.}}
\author[b]{R.~Le~Gac}
\author[e]{J-L.~Meunier}
\author[d]{J-M.~Nappa}
\author[b]{C.~Nassif~Mattar}
\author[f]{C.~Renard}
\author[d]{G.~Vouters}
\affiliation[a]{European Organization for Nuclear Research (CERN),\\Geneva, Switzerland}
\affiliation[b]{Aix Marseille Univ,\\CNRS/IN2P3, CPPM, Marseille, France}
\affiliation[c]{Universit{\'e} Paris-Saclay,\\CNRS/IN2P3, IJCLab, Orsay, France}
\affiliation[d]{Univ. Savoie Mont Blanc,\\CNRS, IN2P3-LAPP, Annecy, France}
\affiliation[e]{LPNHE,\\Sorbonne Universit{\'e}, Universit{\'e} Paris Cit{\'e}, CNRS/IN2P3, Paris, France}
\affiliation[f]{SUBATECH,\\IMT Atlantique, Nantes Universit{\'e}, CNRS-IN2P3, Nantes, France}

\emailAdd{langouet@cppm.in2p3.fr}

\abstract{
The PCIe400 is a generic readout board designed for high-throughput data acquisition in future high-energy physics experiments. It interfaces up to 48 bidirectional links supporting custom protocols from 1 to 26~\gbps to modern back-end systems providing up to 400~\gbps bandwidth. \textcolor{red}{The board is developed as a technological demonstrator for the LHCb Upgrade II, which foresees an aggregated throughput of 200~\tbps. In addition to increased bandwidth, the PCIe400 targets deterministic clock distribution to front-end electronics. At a maximum instantaneous luminosity of $1.5~\times~10^{34}~\aunit{cm}^{-2}\aunit{s}^{-1}$, up to 40 proton-proton interactions per bunch crossing are expected in LHCb Upgrade II. The adoption of 4D tracking detectors with time resolutions down to 20~ps motivates clock distribution with phase determinism below 10~ps peak-to-peak across large-scale systems exceeding 2000 nodes. This paper presents the qualification of the PCIe400 prototype board, focusing on high-bandwidth interfaces, including next-generation QSFP112 links, and phase-deterministic clock distribution.}}


\arxivnumber{2602.01422}

\collaboration[c]{}

\proceeding{Topical Workshop on Electronics for Particle Physics - TWEPP2025\\
  6 - 10 October 2025\\
  Rethymnon, Greece}

\notoctrue

\begin{document}
\maketitle
\flushbottom

\section{\textcolor{red}{Qualification and test infrastructure}}

\textcolor{red}{The qualification of a high-performance readout board requires a flexible test infrastructure capable of configuring peripherals, monitoring system parameters, and automating measurements. A dedicated control and testing framework was developed for the PCIe400 to support the validation of key board features including high-bandwidth interfaces and phase-deterministic clock distribution.}

\subsection{\textcolor{red}{Control and peripheral access}}

\begin{figure}[ht]
    \centering
    
    \includegraphics[width=0.4\textwidth]{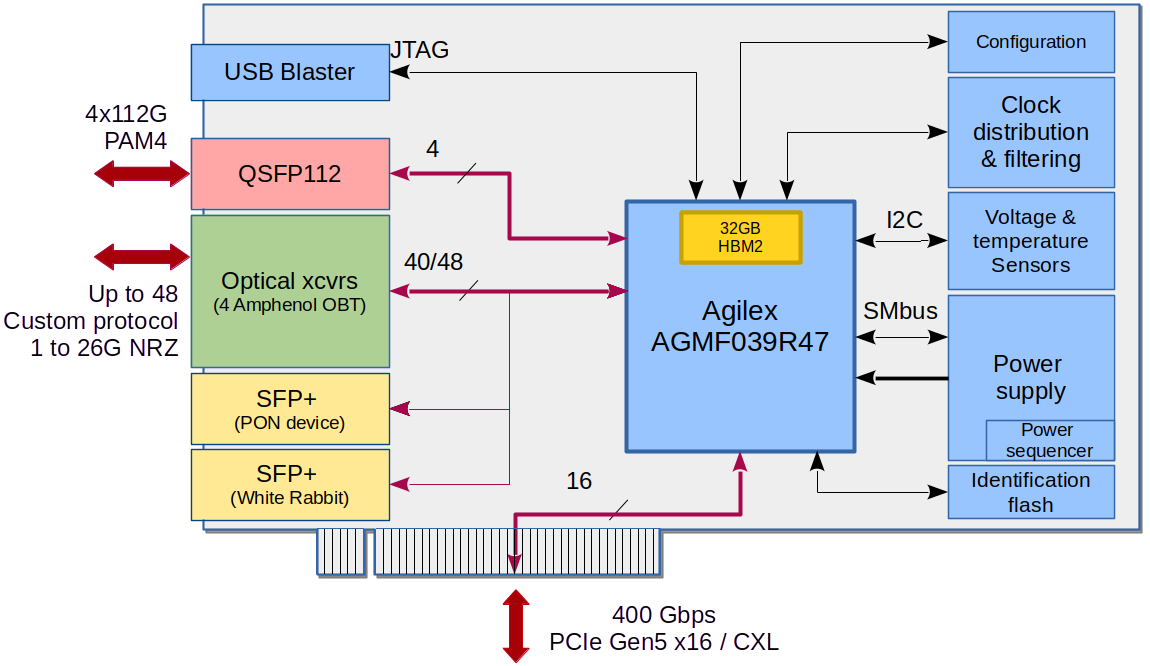}
    \raisebox{0.2\height}{\includegraphics[width=0.42\textwidth]{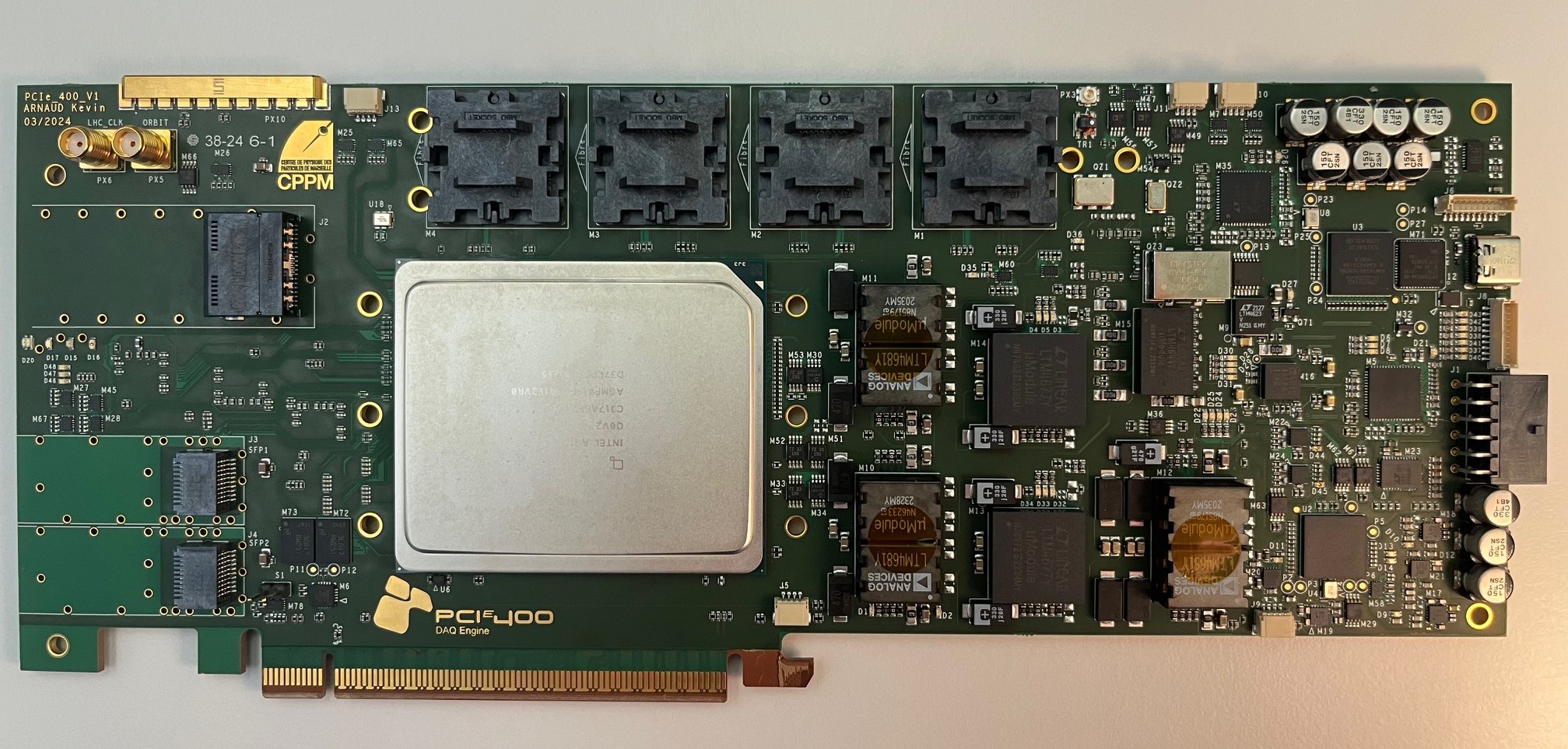}}
    \caption{(left) Block diagram of the PCIe400 board highlighting the FPGA as the central element interfacing PCIe, 400GbE and custom high-speed links. (right) Top view of the PCIe400 prototype illustrating the high component integration density.}
    \label{fig:pcie400_synoptic_photo}
\end{figure}

The board is designed around Altera's Agilex 7 M-series~\cite{altera-agilex} with 1.3~million~ALM\footnote{Adaptive Logic Module} running at up to 1\ghz and 32~GB of HBM2e\footnote{High Bandwidth Memory}. \textcolor{red}{It embeds several peripherals including opto-electronic transceivers, supporting various protocols such as TTC-PON~\cite{9374437} and White~Rabbit~\cite{Serrano:1215571}, an identification flash, power supply monitoring circuits, and jitter filtering PLL.} A block diagram and a photo of the board are given in Figure~\ref{fig:pcie400_synoptic_photo}. 

\textcolor{red}{When installed in a PC server, the board can be controlled through PCIe. A USB/JTAG is also available for debugging and early-stage benchtop validation.} These connections allow the configuration of peripherals and the monitoring of the board status. External components and internal gateware registers are accessible through a common Avalon Memory-Mapped bus.

\subsection{\textcolor{red}{Software framework for qualification tests}}

\begin{figure}[htb] 
    \centering
    \includegraphics[width=0.8\textwidth]{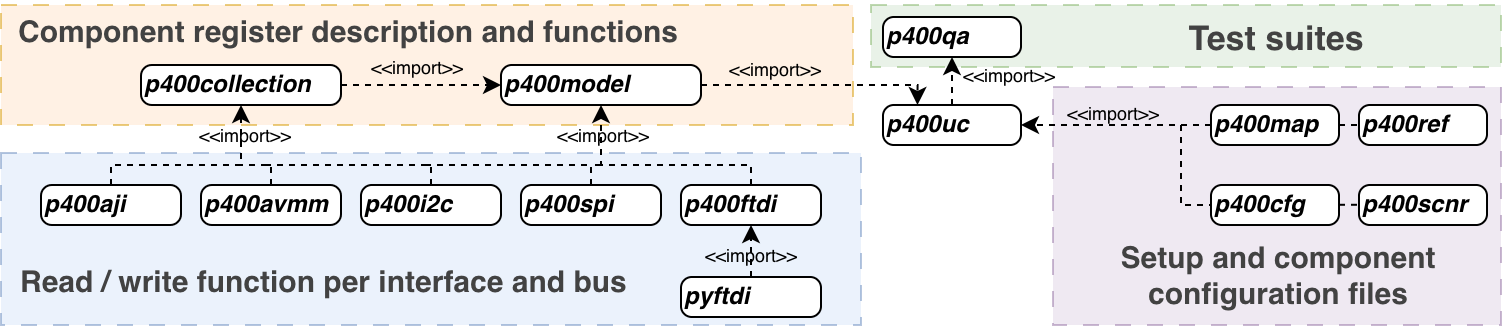}
    \caption{Software architecture of the PCIe400, showing the layered organization used for functional validation, monitoring, and interactive debugging.}
    \label{fig:soft_framework}
\end{figure}

\textcolor{red}{The framework allows the definition of board configuration sets for different test scenarios. It is generic and organized in four layers, with read/write functions for each interface and bus, a collection of register descriptions and methods to interact with each peripheral, a library of configuration files, and a set of tools to execute test suites, as illustrated in Figure~\ref{fig:soft_framework}.}

Three major test suites are implemented: a \textit{Functional Test} comparing sensor values to references, a \textit{Collect Test} to monitor a selection of parameters over time and a \textit{Console} for interactive debugging. The test suites, built upon pytest~\cite{pytest} and polars~\cite{polars}, validate peripheral functionality through more than 400~compliance tests executed in less than a minute.

\section{High bandwidth interface}

\textcolor{red}{In current data acquisition architecture, data is transferred to the host server memory through PCIe and then to the event builder network using commercial network interface cards. A possible evolution of this architecture is to integrate a network interface based on 400GbE directly on the FPGA.} One figure of merit for assessing high-bandwidth interface integrity is the bit error rate (BER). \textcolor{red}{Measurements were performed on both back-end interfaces using pseudo random bit patterns (PRBS).}

\subsection{PCIe interface}

\begin{figure}[htb] 
    \centering
    \includegraphics[width=1.0\textwidth]{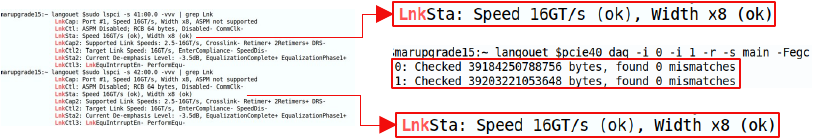}
    \caption{PCIe tests performed on a PCIe Gen~4 server in 2×8 bifurcated mode, demonstrating correct enumeration and DMA transfers with a BER below $1\cdot10^{-15}$ at 95\% confidence level on both interfaces.}
    \label{fig:pcie_test}
\end{figure}

The PCIe interface test is based on previous developments for the generic readout board PCIe40~\cite{cru-firmware}. Thus, the tests were performed in 2x8 bifurcated mode, as data streams could always be split in\textcolor{red}{to} two in previous applications and because it simplifies gateware implementation. \textcolor{red}{The tests are performed on a PCIe Gen~4 server while awaiting access to a Gen~5 server. As shown in Figure~\ref{fig:pcie_test}, correct enumeration of both interfaces is observed and DMA transfers achieving $1\cdot10^{-15}$~BER~$(CL=0.95)$ on both interfaces.}

\subsection{4x100 Gbit/s interface}

The Agilex F-Tile FHT transceiver supports 116\gbps PAM4 per lane and includes a 400GbE hard IP enabling the integration  of a high-bandwidth network interface on-board. The main challenge lies in the \textcolor{red}{physical layer link validation} between the FPGA and the QSFP112 connector, separated by approximately 50\mm~\cite{pcie400-design-challenge}. The test setup uses an FS QSFP112-SR4-400G module in external loopback and Altera’s reference design featuring RS(544,~514) \textcolor{red}{Forward Error Correction} (FEC) with 15~correctable symbols. Seven hours of PRBS testing yields excellent signal integrity \textcolor{red}{with no errors post-FEC and} a maximum of two corrected symbols per codeword.

\section{Phase determinism}

\textcolor{red}{The front-end electronics use the down-link from a back-end board, such as the PCIe400, to recover a reference clock. In order to guarantee a clock distribution with phase determinism below 10~ps peak-to-peak, the transmitter stream must be stable with respect to the LHC master clock across resets, power cycles, and FPGA temperature variations. The study presented here focuses on this first challenge only. In fact,} according to the datasheet, the Agilex FGT transceivers exhibit a non-deterministic lane-to-lane output skew of up to $\pm~2$~unit~interval~(UI)~+~300\ps~\cite{altera-agm-datasheet}. This translates to jumps of 500\ps at 10\gbps and prevents a straightforward solution to distribute phase-deterministic clocks to the front-end. Tests were intentionally performed on an Agilex DK-SI-AGI027FA~development kit in order to isolate the intrinsic phase behavior of the transceiver from board-level effects.~\cite{altera-dk}

\subsection{Principle and set-up}

\begin{wrapfigure}{r}{0.5\textwidth}  
    \includegraphics[width=0.45\textwidth]{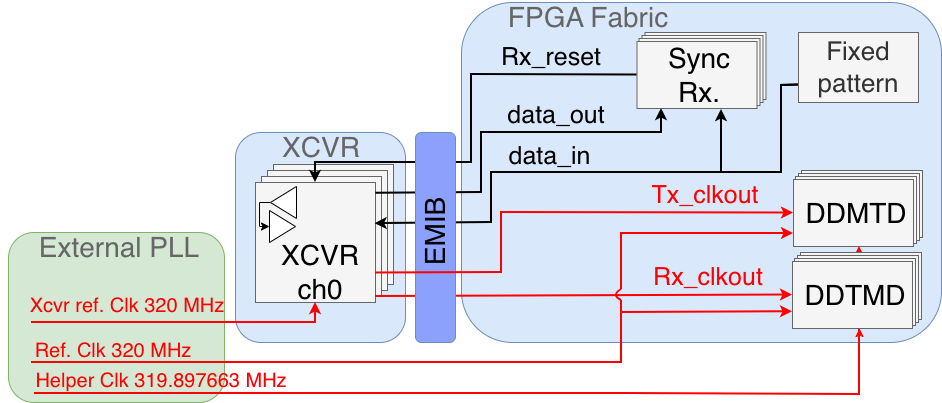}
    \caption{Test setup implemented on the development kit to study transceiver phase behavior. Four serial channels operate at 10.24 Gbit/s, with phase differences measured internally using DDMTD.}
    \label{fig:synoptic_phase}
\end{wrapfigure}

The design implements four channels running at 10.24\gbps. The phase difference with reference clock is measured internally using a 1\ps resolution DDMTD~\cite{ddmtd} as described in Figure~\ref{fig:synoptic_phase}.

\begin{figure}[h] 
    \centering
    \includegraphics[width=0.5\textwidth]{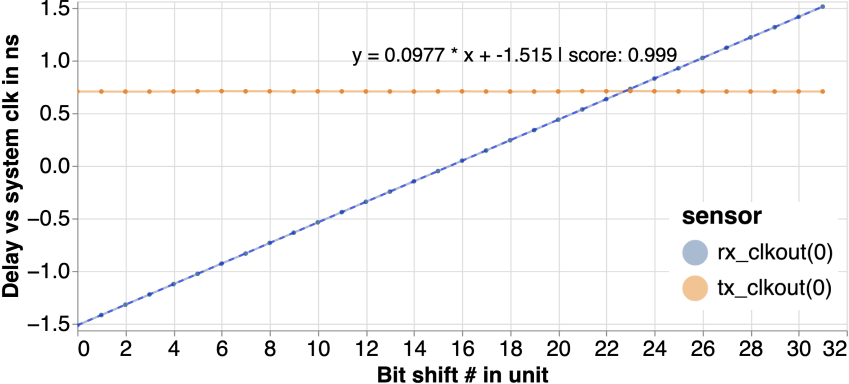}
    \caption{\textcolor{red}{After intentionally shifting the transmitted data pattern by one bit and repeating the receiver synchronization, the phase of the recovered clock is displaced by exactly one unit interval allowing for the monitoring of the transmitter stream phase.}}
    \label{fig:meas1}
\end{figure}

The \textcolor{red}{transmitter clock} cannot be used to measure the phase of the transmitter stream due to non-deterministic internal alignment. \textcolor{red}{Instead, we operate the transceiver in serial-loopback and synchronize the recovered clock to obtain an image of the transmitter stream. Synchronization is achieved by performing receiver resets until the word frontier matches with a fixed 32b~fixed pattern transmitted. Figure~\ref{fig:meas1} illustrates the ability to observe transmitter phase shift using this mechanism.}

\subsection{Tx reset stability}

\begin{figure}[htb] 
    \centering
    \includegraphics[width=0.55\textwidth]{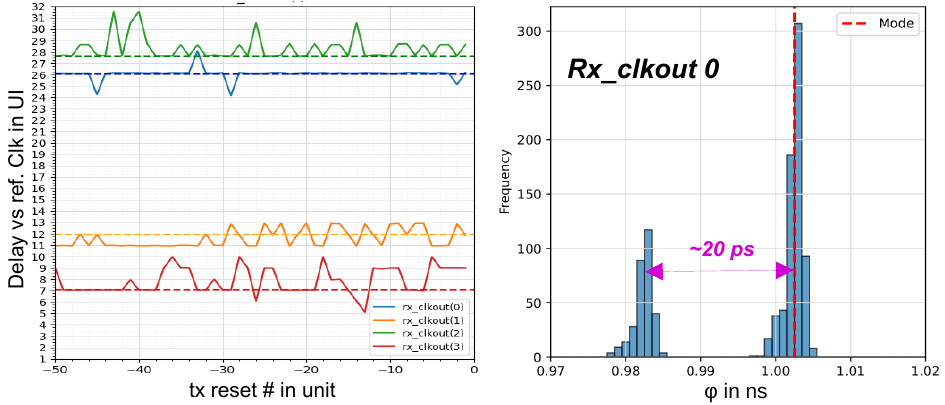}
    \caption{(left) Quantized phase shifts of the recovered clock observed after successive transmitter resets, showing jumps limited to a few unit intervals. (right) Phase distribution obtained from 500 receiver synchronizations without transmitter reset, revealing two stable locking positions of the receiver clock recovery.}
    \label{fig:meas2}
\end{figure}

\textcolor{red}{Now that we have established the ability to measure the transmitter phase shift, we study the effect of transmitter reset.} Figure \ref{fig:meas2} (left) shows that successive transmitter resets produce quantized phase shift of $\pm3\,\mathrm{UI}\allowbreak~(\approx 300~\mathrm{ps})$. With increased statistics, a bimodal distribution appears around each bit position, separated by~$\approx 20\ps$. A test with 500~Rx synchronizations and no Tx reset reveals that this double-peak pattern persists, as in Figure~\ref{fig:meas2}~(right). It indicates that the phase uncertainty results from the receiver path and likely from the Clock Data Recovery circuit locking on two stable positions. Thus, \textcolor{red}{it only impacts the phase monitoring system. This ambiguity is mitigated by selecting the most probable locking position as the reference for the monitoring system.}
%This can be mitigated by the selecting the most probable \textcolor{red}{positionthe monitoring system}.

\begin{figure}[htb] 
    \centering
    \includegraphics[width=0.8\textwidth]{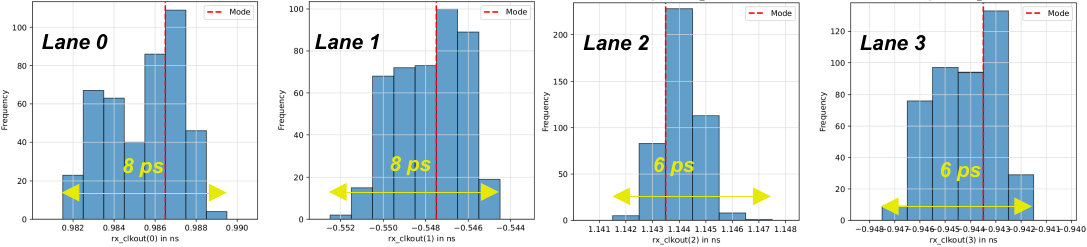}
    \caption{Phase stability of the recovered clock measured over four serial channels after 500 transceiver resets. The distributions show a peak-to-peak phase variation of 6–8 ps.}
    \label{fig:meas4}
\end{figure}

Finally, we assess the inter-reset stability of the transmitter serial stream on four channels. \textcolor{red}{A transmitter reset roulette is applied until the phase is within a $20 \ps$ window centered on previously measured most probable phase position}. The result, in Figure~\ref{fig:meas4}, shows a stability of 6~to~8\ps~peak-to-peak over four~channels after 500~resets during 12~hours demonstrating sub-10 ps stability across channels.

\section{Summary}

In the process of qualifying the generic readout board PCIe400, we developed a versatile software framework that enables efficient testing of complex boards from debugging to production. Extensive simulation work during design allowed the implementation of \textcolor{red}{error-free} 4x100\gbps PAM4 links and PCIe interface. A phase-determinism proof of concept on four channels shows that Agilex FGT transceivers are compatible with a 10\ps~peak-to-peak phase determinism precision \textcolor{red}{between resets}.

Additional tests are underway to qualify all PCIe400 features. In particular, further studies on clock distribution with phase determinism are required to evaluate stability across gateware compilations, temperature variations and at a system level.

\clearpage

\acknowledgments

The authors acknowledge support from CERN and from the IN2P3. We are indebted to the communities behind the multiple open-source software packages on which we depend. We thank our former colleague, Jean-Pierre Cachemiche, for his ingenious and tenacious contribution.



\begin{thebibliography}{99}

\bibitem{altera-agilex}

Altera,
\emph{\href{https://www.intel.com/content/www/us/en/content-details/762901/agilex-7-fpgas-and-socs-product-brief.html}{Agilex 7\texttrademark\xspace  FPGAs and SoCs Product Brief}},
(2024).

\bibitem{9374437}

E.~Mendes, S.~Baron, C.~Soos and F.~Vasey, 
\emph{A Timing, Trigger, and Control System With Picosecond Precision Based on 10 Gbit/s Passive Optical Networks for High-Energy Physics}, 
IEEE Transactions on Nuclear Science, \emph{vol.} 68, no. 4, pp. 447-457 (2021).
%April 2021.
% , doi: 10.1109/TNS.2021.3065128.

\bibitem{Serrano:1215571}

J.~Serrano, P.~Alvarez, M.~Cattin, E.~Garcia~Cota, J.~Lewis, P.~Moreira, T.~Wlostowski, G.~Gaderer, P.~Loschmidt, J.~Dedic, R.~Bär, T.~Fleck, M.~Kreider, C.~Prados and S.~Rauch, \emph{The White Rabbit Project},
CERN-ATS-2009-096,
(2009).

\bibitem{analog-an1144}

Analog Devices,
\emph{\href{https://www.analog.com/media/en/technical-documentation/app-notes/an-1144.pdf}{Measuring Output Ripple and Switching Transients in Switching Regulators}},
(2013).

\bibitem{picmg-pcie5}

PCI-SIG,
\emph{PCI Express Card Electromechanical specification Revision 5.1},
(2023).

\bibitem{altera-cvp}

Altera,
\emph{\href{https://www.intel.com/content/www/us/en/docs/programmable/683763/23-1/fpga-power-supplies-ramp-time-requirement.html}{Agilex 7\texttrademark\xspace Device Configuration via Protocol (CvP) Implementation User Guide}},
(2025).

\bibitem{pytest}

Holger Krekel et al.,
\emph{pytest: simple powerful testing with Python}, 
\href{https://pytest.org/}{https://pytest.org/},
(2015\textendash2025).

\bibitem{polars}

Ritchie Vink et al.,
\emph{Polars: high-performance DataFrame library for Python}, 
\href{https://pola.rs/}{https://pola.rs/},
(2020\textendash2025).

\bibitem{cru-firmware}

O.~Bourrion, J.~Bouvier, F.~Costa, E.~David, J.~Imrek, T.M.~Nguyen, S.~Mukherjee,
\emph{Versatile firmware for the Common Readout Unit
(CRU) of the ALICE experiment at the LHC},
arXiv:1910.08804v2.

\bibitem{pcie400-design-challenge}
P.~Bibron, J.P.~Cachemiche, J.~Langou{\"e}t,
\emph{Technical challenges designing a prototype common readout board for LHCb future upgrades},
arXiv:2411.01648.

\bibitem{altera-agm-datasheet}

Altera,
\emph{\href{https://www.intel.com/content/www/us/en/docs/programmable/769310/current/intel-agilex-7-fpgas-and-socs-device.html}{Agilex\texttrademark\xspace 7 FPGAs and SoCs Device Data Sheet: M-Series}},
(2025).

\bibitem{altera-dk}

Altera,
\emph{\href{https://www.intel.com/content/www/us/en/docs/programmable/721605/current/overview.html}{Agilex\texttrademark\xspace 7 FPGA I-Series Transceiver-SoC Development Kit User Guide}},
(2025).

\bibitem{ddmtd}

P.~Moreira, P.~Alvarez, J.~Serrano, I.~Darwezeh and T.~Wlostowski, 
\emph{Digital dual mixer time difference for sub-nanosecond time synchronization in Ethernet},
IEEE International Frequency Control Symposium, 
(2010).


% Please avoid comments such as "For a review'', "For some examples",
% "and references therein" or move them in the text. In general,
% please leave only references in the bibliography and move all
% accessory text in footnotes.

% Also, please have only one work for each \bibitem.


\end{thebibliography}
\end{document}